\documentclass[aps,prl,twocolumn,showpacs,amsmath]{revtex4}
\newcommand{\meff}{\mbox{$m_{\mbox{\scriptsize eff}}$}}
\newcommand{\phii}{\mbox{$\Phi_{\mbox{\small i}}$}}
\newcommand{\phif}{\mbox{$\Phi_{\mbox{\small f}}$}}
\newcommand{\tif}{\mbox{$T_{\mbox{\small if}}$}}
\newcommand{\phinx}{\mbox{$\phi_{\mbox{\scriptsize n$_x$}}$}}
\newcommand{\phiny}{\mbox{$\phi_{\mbox{\scriptsize n$_y$}}$}}
\begin{document}
%\draft
\preprint{QD-PI/01}
\title{
Oscillations in the photocurrent from quantum dots: Geometric information 
from reciprocal spectra
}
\author{Himadri S. Chakraborty}
\author{Rashid G. Nazmitdinov}
\altaffiliation{Permanent address: Bogoliubov Laboratory of Theoretical Physics,
Joint Institute for Nuclear Research, 141980 Dubna, Russia. }
\author{Mohamed E. Madjet}
\author{Jan-M. Rost}
\affiliation{
Max-Planck-Institut f\"ur Physik Komplexer
Systeme, N\"{o}thnitzer Strasse 38,
D-01187 Dresden, Germany
}
\date{\today}

\begin{abstract}
We demonstrate that 
the current of photoelectrons from a quantum dot 
exhibits oscillations as a function of the photon energy
owing to the
predominant ionization from the dot boundary. In the 
Fourier reciprocal space of the photoelectron wave vector 
these oscillations reveal 
frequencies
connected to the confinement range. 
We attribute the oscillation in the recent photocurrent measurements 
by Fry et al.\ (PRL {\bf 84}, 733 (2000)) to this phenomenon.
Angle-resolved and angle-integrated experiments 
with the focus on directly imaging the 
confining potential 
are suggested. 
\end{abstract}
\maketitle
%PACs descriptions:
%79.60.Jv: Photoemission and photoelectron spectra of nanostructures
%78.67.Hc: Optical properties of quantum dots 
%68.66.Hb: Structural properties of quantum dots  
%\widetext
%\narrowtext

Extraordinary advances in the semiconductor
technology have enabled the fabrication of quantum dots (QDs), 
three-dimensional 
mesoscopic structures,
in which a well-controlled number of electrons can be
confined in a small localized region of space ranging, typically,
from one to several tens of nanometers on a side\cite{bim99,ash96}.
 
QDs reveal features typical for a finite
Fermi system, namely,  
shell structure effects\cite{loc96,tar96,sh},
spin oscillations 
under magnetic field\cite{su98,sp} etc.
However, a common feature in the majority 
of theoretical investigations is to describe the
confinement by a harmonic potential, 
which forbids electron escape.
Such a parabolic description 
works  
for the ground state or the few lowest-lying excited states.
Because, while the depth of the confining potential is 
around 1 eV, the single electron level spacings are 
typically a few meV implying that electrons
in QDs tend to fall in towards the bottom of the potential. 
To these electrons
the top part of the potential remains inaccessible, and therefore, 
whether or not the confining
potential is parabolic, becomes virtually unimportant.
 
On the other hand, this aspect of the confinement
gains importance 
when the study is primarily geared towards understanding 
the interaction dynamics of QDs, for instance, in treating
the photoabsorption induced strong electronic excitations 
followed by radiative decay---the photoluminescence. 
A highly excited
electron feels the top part of the confining potential as well, 
and hence, may become sensitive to the latter's global shape.  
Ultimately, a realistic description of the confining potential 
becomes imminent 
when one considers the photoionization of an electron
from a QD that requires a quantum mechanically legitimate
continuum. 

In the context of photoluminescence experiments on
nanocrystals the indirect
signatures of the photoionization, namely, the
so-called on/off behavior and the spectral diffusion                                  
effect\cite{nir96},
the persistent hole burning\cite{mas95}, photo-darkening\cite{oda00}
etc., have already been observed over the last
years.
A few years ago, shining 476.5 nm laser light on the CdTe nanocrystal
Shen {\em et al}.\ \cite{she99} were able to
detect confirmed evidence
of the photoionization from a QD. 
Very recently, the photocurrent spectroscopy has been employed
to address several structural properties of InAs-GaAs self-assembled
QDs\cite{fry00}.  

Here, we address the photoescape
of an electron from a single QD.  
The photon intensity is assumed to be weak such that the photon interaction 
with electrons 
is predominantly linear to the vector potential 
of the field\cite{brj}. 
We consider for simplicity 
the independent particle description 
of the ionization dynamics and the spin is neglected for
the sake of clarity.  
Further, since in most experiments an electron is tightly bound in a
{\em quantum well} in the $z$-direction, 
we assume a two-dimensional 
confinement in the lateral $xy$-space.
We first derive our results for a finite square well confinement and
then argue that even for a more realistic confinement of electrons
the relevant
feature of the result survives.  
 
We begin with the picture that the absorption of a photon 
of energy $h\nu$ induces the transition of an electron
from an initial state $\phii(x,y)$, bound in the 
QD, to a final 
continuum state $\phif({\bf k};x,y)$ (with ejection wave vector ${\bf k}$)
embedded in the conduction band of the crystal.
The {\em full} (including all polarities) transition amplitude\cite{brj} is 
given by the following two-dimensional integral:
\begin{equation}\eqnum{1}\label{amp1}
\tif = \langle\phif({\bf k};x,y)|{\bf e}\cdot {\bf D}\exp(i{\bf k}_{\nu}
       \cdot {\mathbf \rho})|\phii(x,y)\rangle. 
\end{equation}
Here ${\bf e}$ and ${\bf k}_{\nu}$ are respectively the polarization 
and the wave vector of the photon; ${\mathbf \rho}$ denotes a general position vector
in the $xy$-plane. The  
dipole operator 
${\bf D}$ is $-i{\bf \nabla}_{\rho}$.  

We choose 
${\bf e}$ to be in the $xy$-plane. Now, the photon wave vector
${\bf k}_{\nu}$ must be perpendicular to ${\bf e}$. However,  
the choice of normal incidence, i.e. ${\bf k}_{\nu}$ perpendicular to the $xy$-plane,
renders
$\exp(i{\bf k}_\nu\cdot{\mathbf \rho})$ in Eq.\ (\ref{amp1}) unity.
Equivalently,
only the dipole term of the interaction is retained while contributions
from all higher polarities become {\em identically} zero. The implication
of this choice of the reaction geometry is that while energetic
photons generally induce ionization beyond dipole, this possibility
is entirely preempted in our case.  
It now becomes straight forward to employ 
an {\em equivalent} gauge
for ${\bf D}$ such that  
${\bf D} =
-i/(h\nu){\bf \nabla}_{\rho}V$, with $V$ being the formal 
confining potential.
This acceleration form of the interaction 
embodies the notion that a recoil force $-{\bf \nabla}V$ must
be available to the electron to successfully ionize upon photon
impact; a crucial implication of this mechanism to the present case will be
discussed later.  

The confinement $V(x,y)$ 
is chosen to be a square well of finite height $V_0$:
\begin{equation}\eqnum{3}\label{pot}
V(x,y) = \left\{ 
         \begin{array}{cl}
          V_0 & \mbox{if}~~~~x \ge |a|~~~\mbox{and}~~~y \ge |b|\\
          0 & \mbox{elsewhere}
         \end{array}\right.,
\end{equation}
$2a$ and $2b$ being the extensions of the potential in $x$- 
and $y$-direction respectively.
Since the derivative of a square well potential is a pair of Dirac
$\delta$-functions,  
\begin{equation}\eqnum{4}\label{acc_sim}
{\bf e}\!\cdot\! {\bf D} \!\!=\!\!\frac{V_0}{ih\nu}[e_x\! \{\delta(x\!\!-\!\!a)\!\!-
                     \!\delta(x\!\!+\!\!a)\}
                     \!+\! e_y\! \{\delta(y\!\!-\!\!b)\!\!-\!\delta(y\!\!+\!\!b)\}],
\end{equation}
in which $e_x$ and $e_y$ are the components of
${\bf e}$.
With a symmetric potential (\ref{pot}) the two-dimensional 
Schr\"{o}dinger equation  
is trivially separable and, hence we can formally express $\phii$ as 
\begin{equation}\eqnum{5}\label{ini}
\phii(x,y)=\phinx(x)\phiny(y), 
\end{equation}
wherein $\phinx$ and $\phiny$ are the (normalized) bound
states in one-dimension characterized by the quantum numbers 
$n_x$ and $n_y$ respectively; 
these wavefunctions is either
even or odd in the respective co-ordinate. 

The size-quantized electron makes a transition to the continuum   
of the confining potential which is ``structured"  
by the lattice conduction band.
This photoelectron 
contains a momentum ${\bf p}=\hbar{\bf k}$ whose 
magnitude is uniquely defined by the absorbed photon through 
the energy conservation 
\begin{equation}\eqnum{6}\label{e_con}
h\nu=V_0-E_{\mbox{\footnotesize BE}}+\frac{p^2}{2\meff}, 
\end{equation}
where $E_{\mbox{\footnotesize BE}}$ is the binding energy of the
electron inside the dot and $\meff$ is the effective
electron mass. Ignoring the electron-phonon 
interaction entirely, the final Bloch~state of the electron 
corresponding to its wave vector ${\bf k}$ is 
\begin{equation}\eqnum{7}\label{fin}
\phif({\bf k};x,y)=F(k)U(x,y)\exp(ik_xx+ik_yy).
\end{equation}
In Eq.\ (\ref{fin}), $k_x$ and $k_y$ are the components of 
${\bf k}$ with $k^2=k_x^2+k_y^2$, $F(k)$ is the appropriate
energy normalization 
and $U(x,y)$ is a periodic function with the same periodicity
as the two-dimensional lattice potential. 

Now, under the specified choice of geometry, plugging Eqs.\ (\ref{acc_sim}),  
(\ref{ini}), and (\ref{fin}) in Eq.\ (\ref{amp1}) we obtain 
\begin{equation}\eqnum{8}\label{amp2}
\tif=F(k)/(ih\nu)\left[e_xJ_1^a+e_yJ_2^b\right]
\end{equation} 
where
\begin{equation}
J_1^a\!\!\! =\!\!\! \int_{-\infty}^{\infty}\!\!\!\!\!\!\!\!dx
   [\delta(x\!-\!a)\!-\!\delta(x\!+\!a)]
   \phinx\!\exp(\!-ik_xx)W(k_y,x)
\eqnum{9}\label{j1}
\end{equation}
with
\begin{equation}\eqnum{10}\label{w}
W(k_y,x)=V_0\int_{-\infty}^{\infty}dy{\phiny(y)U(x,y)}
     \exp(-ik_yy).
\end{equation}
$J_2^b$ in Eq.\ (\ref{amp2}) is an integral exactly 
as (\ref{j1}) but with the variable $x$ and the 
subscript $x$ being replaced by those of $y$, and {\em vice versa}. 

Let us consider the integral $J_1^a$. 
$W(k_y,x)$ is finite and can in principle be evaluated as a decaying
function of $k_y$. 
If we now assume for simplicity that the periodic function
$U(x,y)$ is even in $x$ then $W(k_y,x)$ also becomes even in $x$.   
Assuming further that
$\phinx$ is odd in $x$ we have the final expression for $J_1^a$: 
\begin{subequations}
\begin{equation}\eqnum{11a}\label{j1_an}
J_1^a=2\phinx(a)W(k_y,a)\cos(ak_x).
\end{equation}
Similarly, assuming $U(x,y)$ and $\phiny$ to be respectively even and odd in $y$,
we obtain
\begin{equation}\eqnum{11b}\label{j2_an}
J_2^b=2\phiny(b)W(k_x,b)\cos(bk_y).
\end{equation}
\end{subequations}
Substituting Eqs.\ (\ref{j1_an}) and (\ref{j2_an}) in 
Eq.\ (\ref{amp2}) we arrive at the final form of 
the transition amplitude,
\begin{equation}\eqnum{12}\label{amp3}
\tif(k_x,k_y)=2F(k)/(ih\nu)[A\cos(ak_x)+B\cos(bk_y)]
\end{equation} 
where the expressions for the $k$-dependent coefficients are
$A=e_x\phinx(a)W(k_y,a)$ and $B=e_y\phiny(b)W(k_x,b)$.
It should, however, be trivially understood that for any other
symmetry combinations of $\phinx$, $\phiny$ and 
$U(x,y)$ we get similar results
with only either or both of the cosine-oscillations in Eq.\ (\ref{amp3})
being replaced by the corresponding sine-oscillations.
Evidently, the amplitude $\tif$ exhibits two oscillations
with frequencies $a$ and $b$ respectively in $k_x$- and $k_y$-space.

Now, the quantum interference effect 
must yield four oscillations in the resulting photoelectron {\em angular 
distribution} spectrum.
Expressing $k_x=k\cos\theta$ and $k_y=k\sin\theta$, where  
${\bf k}$ makes an angle $\theta$ 
with the $x$-axis, the angular distribution 
can be expressed as 
\begin{eqnarray}
\frac{d\sigma}{d\theta}(k)&=&\frac{e^2h^2}
          {2\pi m_{\mbox{\footnotesize eff}}^2c\nu}
          \left|\tif(k_x,k_y)\right|^2\nonumber\\
   &=& \frac{2e^2h^2}{\pi m_{\mbox{\footnotesize eff}}^2c\nu^3}
       F^2(k)
       \left[\frac{A^2}{2}\cos((2a\cos\theta)k)\right.
       \nonumber \\
   && \left.\! +\frac{B^2}{2}
      \!\cos((\!2b\sin\theta)k)\!\!+\!\!AB\!\cos((\!a\cos\theta\!\!+\!\!b\sin\theta)k)\right.
      \nonumber \\
   &&  \left.\! +AB\cos((a\cos\theta\!-\!b\sin\theta)k)\!+\!\frac{A^2}{2}
       \!+\!\frac{B^2}{2}\right],
\eqnum{13}\label{ddx}\end{eqnarray}
which delineates the frequencies, 
$2a\cos\theta$, $2b\sin\theta$, $a\cos\theta+b\sin\theta$,
and $|a\cos\theta-b\sin\theta|$ in the photoelectron
$k$-space.  
Since photoelectrons emanated at an angle $\theta$ generate 
a current in that direction 
through the matrix, the behavior   
of this current as a function of the photon energy 
will also show identical oscillations.  

The finite square well is certainly an idealistic choice to
describe electron trapping in a QD. 
Even if we assume that the external 
confinement induced by abrupt heterojunction interfaces is of nearly
hard-wall nature, the inclusion of electron-electron interactions
must soften the edge of the effective confinement.  
Furthermore, while in many ways the structural properties of QDs
resemble those of natural atoms, they are radically dissimilar
in one particular aspect. For atomic
systems electrons are practically localized
around the nucleus to yield a generic near-Coulomb potential
with a steep slope close to the origin. 
In the context of dots, however, 
electrons in
the interior region of a crystal
are {\em quasi-free} and only sense the potential
well of the confinement.
As a result, the effective
potential 
may be relatively 
flat within the dot but sharply rising at the dot boundary 
to a finite height with a tail representing the residual
Coulomb interaction.
Recent experimental 
evidence\cite{fuh01} of steep-wall confinement in QDs  
strongly supports this assumption.    
Consequently, the gradient of the potential
will show strong peaks (instead of $\delta$-functions) 
near the boundary
while being relatively weak across the interior region. 
This results in a predominant contribution to the relevant overlap
integral in Eq.\ (\ref{amp1}) from the respective peak position. 
Equivalently, the probability that an electron receives sufficient recoil 
to ionize is rather
high near the edge---a fact which is uncovered here through 
the acceleration framework of the electron-photon interaction.
Therefore, even with a realistic confining potential
the photoelectron angular distribution must show
the same oscillatory behavior as in Eq.\ (\ref{ddx}).
The energy-dependence of the background strength of 
electron intensity at an angle
$\theta$ will, of course, be different from the square well case owing
primarily to (i) the more realistic ground state wavefunction 
and (ii) a finite width 
of the peak in the potential gradient. 

Eq.\ (\ref{ddx}) can be integrated over $\theta$ to obtain
the total cross section, in terms of the Bessel function of order zero,
which for large enough $k$ will assume the form:
\begin{eqnarray}
\sigma(k)&\!\simeq\!& \frac{2e^2}{\pi m_{\mbox{\footnotesize eff}}^2c\nu^3}
       \!F^2\!\!\left[\!\frac{A^2}{(ak)^{1/2}}\!\sin(2ak)\!+\!
       \frac{B^2}{(bk)^{1/2}}\!\sin(2bk)\right.
       \nonumber \\
       &&\left. +\frac{4\sqrt{2}AB}{\sqrt{a^2+b^2}k}\sin(\sqrt{a^2+b^2}k)
       \right].
\eqnum{14}\label{tx}\end{eqnarray}
If the potential is circular ($a=b$) and the photoelectron
emission is isotropic ($k_x=k_y$), then Eq.\ (\ref{ddx})
(and hence Eq.\ (\ref{tx})) 
simplifies to an oscillation with a single frequency
being the diameter of the potential.  

Let us now consider Ref.\ \cite{fry00}, in which the measured 
photocurrent spectra from
the self-assembled InAS-GaAs QD are clearly exhibiting
oscillations superimposed on a background. We attribute this feature 
to the mechanism described above. These lens-shaped 
dots of circular base have been illuminated by  
the light, that presumably incidents normally on the surface of the sample. 
This implies that the resulting oscillation should correspond to a 
frequency connected to the diameter
of a circle, which is the locus of the peak of the
derivative of the two-dimensional lateral potential that confines
the electron. As further evident in Ref.\ \cite{fry00},
with increasing photon energy ($h\nu$) the period of the 
oscillation increases but remains constant
as a function of $(h\nu)^{1/2}\propto k$. 
This is exactly what we expect from the 
result derived here that the spectra oscillate with
equi-distant extrema in the photoelectron $k$-space. A rough estimation
of this frequency $2\pi/\Delta k$, with $\Delta k$ being
the period length of the oscillation
in the corresponding $k$-space, 
yields a diameter of about
50 nm using $\meff$ for InAs to be about 3\% of the free electron
mass. Furthermore, 
raising each of the applied bias voltage and the dot temperature  
will simultaneously increase the  
energy of the confined electrons and decrease
the effective potential height.   
As a result, ionization occurs at
progressively lower photon energies (see Eq.\ (\ref{e_con})). 
This qualitatively  
explains why the spectrum in Ref.\ \cite{fry00}
suffers a constant shift to the low energy side
as a result of increasing bias voltage and temperature.  
For a quantitative analysis detailed calculation will be
published elsewhere. 
We comment here that oscillation from a similar phenomenology
has been discussed for
metal clusters 
in the spherical geometry\cite{ola99,mad01,cha01}
and observed for fullerenes\cite{bec00}.      

At this stage we can envisage a 
photoionization experiment on a QD with non-circular
lateral shape 
in order to image its confining potential. 
The sample embedded in a suitable matrix can be exposed to 
a tunable monochromated light which incidents perpendicular to 
the lateral plane. 
We suggest
simultaneous measurements of the photocurrent along two {\em mutually
perpendicular} directions on the lateral plane over
a range of photon energy. 
Such angle-resolved photocurrent measurements can possibly be
conducted by employing electric point-contacts with the 
voltage across the circuit low enough to minimize the leakage current.
Both spectra can then be
transformed to the $k$-space using Eq.\ (\ref{e_con}). In doing so, 
a rough dot-specific estimation
may be used for 
$E_{\mbox{\footnotesize BE}}$---which is a good approximation
since typically $V_0\gg E_{\mbox{\footnotesize BE}}$.
A set of four frequencies for each of the spectra can now be
determined from 
their Fourier transforms after subtracting the steady 
background part of the photocurrent.
Identifying 
$f_1=2a\cos\theta$ and $g_1=2b\sin\theta$ from the first set
and $f_2=-2a\sin\theta$ and
$g_2=2b\cos\theta$ from the second, where $\theta$
is the angle made by the corresponding ${\bf k}$ with the $x$-axis, 
the extensions of the potential in the $x$- and 
$y$- direction can be obtained via 
$2a=\sqrt{f_1^2+f_2^2}$ and $2b=\sqrt{g_1^2+g_2^2}$.
It should also be easy to extract $\theta$ from the frequencies
and thereby determining the angular orientation of the dot. 

If such angle-resolved measurements prove difficult, one can
measure the total photocurrent relatively conveniently
from a set of isolated dots in the similar
manner as in Ref.\ \cite{fry00}. But for
a reliable Fourier signal it may be necessary to extend the data 
over larger energy range. While for circular dots a single peak will
appear in the Fourier spectrum, for non-circular ones additional
peaks will emerge (Eq.\ (\ref{tx})). The geometric information on the 
confinement, hence obtained through the transformation of the real space
photo-signal to its reciprocal Fourier space, 
can now be readily used
to solve the Schr\"{o}dinger equation self-consistently
in order to determine the confining potential, and thereby
the wavefunctions. 
  
For the ionization to
occur the incident photon should have energy greater than $V_0-
E_{\mbox{\footnotesize BE}}$ (see Eq.\ (\ref{e_con})). 
Since 
$E_{\mbox{\footnotesize BE}}$ is 
typically of the order of several meV the influence of
electron collective
effects  
on the cross section
may remain operative only over a small range
above the ionization threshold. Beyond this range the independent
particle model should well explain the ionization phenomenon. 
Therefore, 
a rather safe starting point for the photon energy
to conduct an experiment will be about a hundred meV
above $V_0$. 
One {\em caveat} should be mentioned, however. 
If the photon wavelength reaches a point when 
the matrix begins to absorb it then the signal 
may be contaminated by the photocurrent 
from the matrix itself. 

In recent years, the resonant magnetotunneling 
spectroscopy has been employed as a convenient tool to directly
image the electronic wavefunctions in self-assembled QDs\cite{vdo00}.
Our work shows, on the other hand,
that photoionization in the Fourier space
can serve as an efficient technique to extract valuable knowledge
about the dot geometry. The above statement becomes particularly underscored
due to the desirability of the photoionization process for
this purpose by virtue of its 
nearly non-destructive nature because of the weak coupling between
photons and target electrons. 

To conclude, we have shown that owing to (a) the finite height
of the quantum confinement and (b) the quasi-free delocalized character
of the interior electrons, the photocurrent spectra from a 
QD show oscillations whose frequencies in the
photoelectron $k$-space are connected to the 
lateral extensions of the dot. This feature is 
a direct consequence of the 
mechanism that photoelectrons are predominantly ejected
from the dot boundary. We have identified the oscillatory
behavior of the recent measurements on InAs-GaAs dots with
this predicted effect. Finally, we have suggested 
possible photoionization experiments for measuring the 
geometric extensions and orientation
of the electron confinement in a dot.


\begin{thebibliography}{99}
\bibitem{bim99} D. Bimberg, M. Grundmann, and M.N. Ledentsov,
                {\it Quantum Dot Heterostructures} (John Wiley \&
                Sons, New York, 1998).
\bibitem{ash96} R.C. Ashoori,
                Nature (London) {\bf 379}, 413 (1996).
\bibitem{loc96} D.J. Lockwood et al., 
%               P. Hawrylak, P.D. Wang, C.M. Sotomayor Torres,
%               A. Pinczuk, and B.S. Dennis,
                Phys.\ Rev.\ Lett. {\bf 77}, 354 (1996).
\bibitem{tar96} S. Tarucha et al.,
%               D.G. Austing, T. Honda, R.J. van der Hage,
%               and L.P. Kouwenhoven,
                Phys.\ Rev.\ Lett. {\bf 77}, 3613 (1996).
\bibitem{sh}    W.D. Heiss and R.G. Nazmitdinov,
                Phys.\ Lett.\ A {\bf 222}, 309 (1996);
                Phys.\ Rev.\ B {\bf 55}, 16310 (1997).
\bibitem{su98}  Bo Su, V.J. Goldman, and J.E. Cunningham,
                Phys.\ Rev.\ B {\bf 46}, 7644 (1992).
\bibitem{sp}    M. Dineykhan and R.G. Nazmitdinov,
                Phys.\ Rev.\ B {\bf 55}, 13707 (1996).
\bibitem{nir96} M. Nirmal et al., 
%               B.O. Dabbousi, M.G. Bawendi, J.J. Macklin,
%               J.K. Trauman, T.D. Harris, and L.E. Brus, 
                Nature (London) {\bf 383}, 802 (1996);
                S.A. Empedocles, D.J. Norris, and M.G. Bawendi,
                Phys.\ Rev.\ Lett. {\bf 77}, 3873 (1996).
\bibitem{mas95} Y. Masumoto et al., 
%               S. Okamoto, T. Yamamoto, and T. Kawazoe,
                Phys.\ Status Solidi (b) {\bf 188}, 209 (1995).
\bibitem{oda00} M. Oda et al., 
%               M.Y. Shen, M. Saito, and T. Goto, 
                J. Lumin. {\bf 87}, 469 (2000).
\bibitem{she99} M.Y. Shen, M. Oda, and T. Goto,
                Phys.\ Rev.\ Lett. {\bf 82}, 3915 (1999).
\bibitem{fry00} P.W. Fry et al., 
%               I.E. Itskevich, D.J. Mowbray, M.S. Skolnick, 
%               J.J. Finley, J.A. Barker, E.P. O'Reilly, L.R. Wilson,
%               I.A. Larkin, P.A. Maksym, M. Hopkinson, M. Al-Khafaji,
%               J.P.R. David, A.G. Cullis, G. Hill, and J.C. Clark,
                Phys.\ Rev.\ Lett. {\bf 84}, 733 (2000); 
                Physica E {\bf 7}, 408 (2000).
\bibitem{brj}   B.R. Bransden and C.J. Joachain,
                {\it Physics of Atoms and Molecules} (Longman, New York, 1997).
\bibitem{fuh01} A. Fuhrer et al., 
%               S. L\"{u}scher, T. Heinzel, K. Ensslin, W. Wegscheider,
%               and M. Bichler,
                Phys.\ Rev.\ B {\bf 63}, 125309 (2001).
\bibitem{ola99} Olaf Frank and Jan M. Rost, 
                Phys.\ Rev.\ A {\bf 60}, 392 (1999); 
                Z. Phys.\ D {\bf 38}, 59 (1996). 
\bibitem{mad01} M.E. Madjet, Himadri S. Chakraborty, and Jan-M. Rost,
                J. Phys.\ B {\bf 34}, L345 (2001).
\bibitem{cha01} Himadri S. Chakraborty and M.E. Madjet, 
                LANL Eprint: arXiv:physics/0105050 (2001). 
\bibitem{bec00} U. Becker, O. Gessner, and A. R\"{u}del, 
                J. Elec.\ Spect.\ Rel.\ Phen. {\bf 108}, 189 (2000).
\bibitem{vdo00} E.E. Vdovin et al., 
%               A. Levin, A. Patane, L. Eaves, P.C. Main,
%               Yu.N. Khanin, Yu.V. Dubrovskii, M. Henini, and G. Hill,
                Science {\bf 290}, 122 (2000) 
\end{thebibliography}
\end{document}